\documentstyle[12pt]{article}
         \makeatletter
         \def\thefigure{\@arabic\c@figure}\def\fps@figure{tbp}
         \def\ftype@figure{1}\def\ext@figure{lof}
         \def\fnum@figure{\protect\footnotesize Fig.\ \thefigure}
         \def\thetable{\@arabic\c@table}
         \def\fps@table{tbp}\def\ftype@table{2}\def\ext@table{lot}
         \def\fnum@table{\protect\footnotesize Table \thetable}
         \def\@listI{\leftmargin\leftmargini\parsep=0pt\itemsep=0pt}
         \def\thebibliography#1{\section{References}\vspace*{-10pt}\list
          {[\arabic{enumi}]}{\settowidth\labelwidth{[#1]}\leftmargin\labelwidth
          \advance\leftmargin\labelsep
          \usecounter{enumi}}
          \def\newblock{\hskip .11em plus .33em minus .07em}
          \sloppy\clubpenalty4000\widowpenalty4000
          \sfcode`\.=1000\relax}
         
         \def\@nomath#1{\ifmmode \fi}
         \def\mmycite{\@ifnextchar [{\@tempswatrue\@mmycitex}
             {\@tempswafalse\@mmycitex[]}}
         \def\@mmycitex[#1]#2{\if@filesw\immediate%
         \write\@auxout{\string\citation{#2}}\fi
           \def\@citea{}\@mmycite{\@for\@citeb:=#2\do
             {\@citea\def\@citea{,}\@ifundefined
                {b@\@citeb}{{\bf ?}\@warning
                {Citation `\@citeb' on page \thepage \space undefined}}%
         \hbox{\csname b@\@citeb\endcsname}}}{#1}}
         \def\@mmycite#1#2{{{\scriptsize#1}\if@tempswa , #2\fi}}
         \def\mycite#1{$^{\protect\mmycite{#1}}$}
         \def\@cite#1#2{{#1\if@tempswa , #2\fi}}
         \def\thesection {\arabic{section}}
         \def\section#1{\addtocounter{section}{1}\setcounter{subsection}{0}
              \bigskip\medskip{\noindent\bf\thesection.\ #1}
              \medskip}
         \def\thesubsection {\arabic{section}.\arabic{subsection}}
         \def\subsection#1{\addtocounter{subsection}{1}
              \medskip{\noindent\thesubsection.\ #1}
              \medskip}
         \makeatother

\def\au#1 {\begin{center} #1 \end{center}}
\def\case#1/#2{{\textstyle\frac{#1}{#2}}}
\def\cntr#1 {\begin{center} #1 \end{center}}
\def\eq#1 #2 {\begin{equation} \label{#1} #2 \end{equation}}
\def\eqa#1 #2 #3 {\begin{eqnarray} \label{#1} #2 \label{#3} \end{eqnarray}}

\def\fig#1 #2 #3 #4 {\begin{figure} \vspace{#3pt} \caption[#1]{#4} \label{#1}
\end{figure}}

\def\tbl#1 #2 #3 #4 {\begin{table} \caption[#1]{#3} \label{#1} \vspace{-6pt}
\begin{center} {\begin{tabular}{#2}  \hline\hline #4 \vspace{1pt} \\
\hline\hline \end{tabular}} \end{center} \end{table}}

\def\tblb#1 #2 #3 #4 {\begin{table}[b] \caption[#1]{#3} \label{#1}
\vspace{-6pt} \begin{center} {\begin{tabular}{#2}  \hline\hline #4 \vspace{1pt}
\\ \hline\hline \end{tabular}} \end{center} \end{table}}

\def\tblh#1 #2 #3 #4 {\begin{table}[h] \caption[#1]{#3} \label{#1}
\vspace{-6pt} \begin{center} {\begin{tabular}{#2}  \hline\hline #4 \vspace{1pt}
\\ \hline\hline \end{tabular}} \end{center} \end{table}}

\def\ti#1 {\begin{center} \baselineskip=17pt {\large #1} \end{center}}

\def\tibf#1 {\begin{center} \baselineskip=17.5pt {\large \bf #1} \end{center}}

\def\aleq{\hspace{-6.712pt}&=&\hspace{-6.712pt}}
\def\alsp{\hspace{-6.712pt}& &\hspace{-6.712pt}}

\def\ess{\hskip.444444em plus .499997em minus .037036em}
\def\mss{\hskip.333333em plus .208331em minus .088889em}

\def\sen{\hbox{\scriptsize--}}
\def\eV{e\kern-.10emV }
\def\eVcm{e\kern-.10emV\kern-.15em,\mss}
\def\eVcl{e\kern-.10emV\kern-.10em:\ess}
\def\eVsc{e\kern-.10emV\kern-.10em;\mss}
\def\eVp{e\kern-.10emV\kern-.15em.\ess}
\def\eVpr{e\kern-.10emV) }
\def\eVc{e\kern-.10emV\kern-.10em/\kern-.10em$c$ }
\def\eVccm{e\kern-.10emV\kern-.10em/\kern-.10em$c$, }
\def\eVcp{e\kern-.10emV\kern-.10em/\kern-.10em$c$. }
\def\eVf{e\kern-.10emV\kern-.10em/fm }
\def\eVfcm{e\kern-.10emV\kern-.10em/fm, }
\def\eVfp{e\kern-.10emV\kern-.10em/fm. }
\def\bit{\begin{itemize}}
\def\eit{\end{itemize}}
\def\qids{\dot{q}_i^{\;2}}

\def\gs{{g_{\rm s}}}
\def\gv{{g_{\rm v}}}
\def\ms{{m_{\rm s}}}
\def\mv{{m_{\rm v}}}
\def\msv{{m_{\rm s,v}}}

\def\part{{\partial_t}}

\begin{document}

\vspace*{16pt}
\au{CLASSICAL HADRODYNAMICS\@: A NEW APPROACH\\
TO ULTRARELATIVISTIC HEAVY-ION COLLISIONS}
\vspace{6.1pt}

\au{Brian W. Bush and \underline{J. Rayford Nix}\\
{\it Theoretical Division, Los Alamos National Laboratory \\
Los Alamos, New Mexico 87545, USA}}
\vspace{7.0pt}

\begin{abstract}
We discuss a new approach to ultrarelativistic heavy-ion collisions based on
classical hadrodynamics for extended nucleons, corresponding to nucleons of
finite size interacting with massive meson fields.  This new theory provides a
natural covariant microscopic approach that includes automatically spacetime
nonlocality and retardation, nonequilibrium phenomena, interactions among all
nucleons and particle production.  In the current version of our theory, we
consider $N$ extended unexcited nucleons interacting with massive neutral
scalar ($\sigma$) and neutral vector ($\omega$) meson fields.  The resulting
classical relativistic many-body equations of motion are solved numerically
without further approximation for soft nucleon-nucleon collisions at \linebreak
$p_{\rm lab}$ = 14.6, 30, 60, 100 and 200 G\eVc to yield the transverse
momentum imparted to the nucleons.  For the future development of the theory,
the isovector pseudoscalar ($\pi^+$,~$\pi^-$,~$\pi^0$), isovector scalar
($\delta^+$,~$\delta^-$,~$\delta^0$), isovector vector
($\rho^+$,~$\rho^-$,~$\rho^0$) and neutral pseudoscalar ($\eta$) meson fields
that are known to be important from nucleon-nucleon scattering experiments
should be incorporated.  In addition, the effects of quantum uncertainty on the
equations of motion should be included by use of techniques analogous to those
used by Moniz and Sharp for nonrelativistic quantum electrodynamics.
\end{abstract}

\section{Introduction}

Many of you are involved in the search for the quark-gluon plasma---a~predicted
new phase of nuclear matter where quarks roam almost freely throughout the
medium instead of being confined to individual nucleons.  Experimental
identification of the quark-gluon plasma will require accurate predictions for
ultrarelativistic heavy-ion collisions on the basis of conventional nuclear
physics, with the known hadronic degrees of freedom {\it properly\/} taken into
account.  Significant deviations between these predictions and experimental
data would then signal the onset of new phenomena such as a quark-gluon
plasma.

With this goal in mind, we have developed at Los Alamos an entirely new
microscopic many-body approach to ultrarelativistic heavy-ion collisions based
on classical hadrodynamics for extended nucleons, corresponding to nucleons of
finite size interacting with massive meson fields.  This approach, which
satisfies {\it a priori\/} the physical conditions that exist at
ultrarelativistic energies, is manifestly Lorentz-covariant and allows for
nonequilibrium phenomena, interactions among all nucleons and particle
production.

The physical input underlying this approach consists of Lorentz invariance
(which includes energy and momentum conservation), nucleons of finite size
interacting with massive meson fields and the classical approximation applied
in domains where it should be reasonably valid.  This starting point builds
upon the traditional hadronic description of nuclear processes that has in the
past been so successful in a wide variety of situations.  By treating the
underlying quarks and gluons implicitly in terms of nucleons and mesons, we are
able to solve the resulting classical relativistic many-body equations of
motion numerically without further approximation.  In particular, we do not
need to make either a mean-field approximation, a perturbative expansion in
coupling strength or a superposition of two-body collisions.

The motivation for the classical approximation in this new theory is that at
bombarding energies of many G\eV per nucleon, the de~Broglie wavelength of
projectile nucleons is extremely small compared to all other length scales in
the problem.  In addition, the Compton wavelength of the nucleon is small
compared to its radius, so that effects due to the intrinsic size of the
nucleon dominate those due to quantum uncertainty.  Finally, the angular
momentum is typically several hundred $\hbar$ and the radiated energy
corresponds to several meson masses.  The classical approximation for nucleon
trajectories should therefore be valid, provided that the effect of the finite
nucleon size on the equations of motion is taken into account.

We describe in sect.~2 the present version of our theory, which includes the
neutral scalar ($\sigma$) and neutral vector ($\omega$) meson fields.  This
permits a qualitative discussion of such physically relevant points as the
effect of the finite nucleon size on the equations of motion, an inherent
spacetime nonlocality that may be responsible for significant collective
effects and particle production through massive bremsstrahlung.  The $\sigma$
and $\omega$ mesons that are produced through this mechanism will subsequently
decay primarily into pions with some photons also.  The resulting classical
relativistic equations of motion are solved in sect.~3 for soft nucleon-nucleon
collisions at $p_{\rm lab}$ = 14.6, 30, 60, 100 and 200 G\eVc to yield the
transverse momentum imparted to the nucleons.  Section~4 discusses future
directions for the systematic development of the theory.  Further details are
given in a series of papers,\mycite{SHN90}$^{\sen}$\mycite{BN94}\mss although
not all of the equations appearing in some of the earlier publications are in
their final form.
\bigskip

\section{Equations of motion}

Our action for $N$ extended unexcited nucleons interacting with massive scalar
and vector meson fields is
\begin{eqnarray}\label{action}
I \aleq \overbrace{- M_0\sum_{i=1}^N \int d \tau_i \, \sqrt{\qids}}^{\sf
Nucleons} + \overbrace{\frac{1}{8\pi} \int d^4 \! x \left [(\partial \phi)^2 -
\ms^{\!2} \, \phi^2 \right ]}^{\sf Scalar~field} \nonumber \\ \alsp \mbox{}-
\underbrace{\frac{1}{8\pi} \int d^4 \! x \left ( \frac{1}{2} G^2 - \mv^{\!2} \,
V^2 \right )}_{\sf Vector~field} - \underbrace{\int d^4 \! x \left ( j \phi + K
\cdot V \right )}_{\sf Interaction} ~,
\end{eqnarray}
where $M_0$ is the bare nucleon mass and $q_i$ is the four-position of the
$i$th nucleon, whose trajectory is given by $q_i = q_i(\tau_i)$.  A dot
represents the derivative with respect to $\tau_i$.  In the action the
four-velocities are not constrained so that $\qids = 1$ and $\tau_i$ is not yet
identified as the proper time; it is only in the equations of motion, which are
derived as a result of the variation of $I$, that this is true.  We use the
metric $g^{\mu\nu} = {\rm diag}(1,-1,-1,-1)$, write four-vectors as $q^\mu =
(q^0, {\bf q}) = (q^t, q^x, q^y, q^z)$ and use units in which $\hbar = c = 1$.
The scalar potential is denoted by $\phi$, the four-vector potential by $V$ and
the meson masses by $\msv$.  The vector field strength tensor is
\begin{equation}\label{vfst}
G^{\mu\nu}  =  \partial^\mu V^\nu - \partial^\nu V^\mu ~,
\end{equation}
the scalar source density is
\begin{equation}\label{ss}
j(x)  = \gs \sum_{i=1}^N \int d \tau_i \, \rho(x - q_i, \dot{q}_i) \sqrt{\qids}
\end{equation}
and the vector source density is
\begin{equation}\label{vs}
K^\mu(x) = \gv  \sum_{i=1}^N \int d \tau_i \, \rho(x - q_i, \dot{q}_i) \,
\dot{q}_i^\mu ~,
\end{equation}
where $\rho$ is the four-dimensional mass density of the nucleon, the spatial
part of which we assume to be exponential in the nucleon's rest frame.  The
values\mycite{SW86}$^{\sen}$\mycite{So79} that we are currently using for the
six physical constants appearing in our theory are nucleon mass \linebreak $M$
$=$ 938.91897 M\eVcm scalar ($\sigma$) meson mass $\ms$ $=$ 550 M\eVcm vector
($\omega$) meson mass $\mv$ $=$ 781.95 M\eVcm scalar interaction strength
$\gs^2$ $=$ 7.29, vector interaction strength \linebreak $\gv^2$ $=$ 10.81 and
nucleon r.m.s.\ radius $R_{\rm rms}$ $=$ 0.862 fm.

In ref.~\cite{BN93} we have derived exact equations of motion for the above
action in two limits:  (1) relativistic point nucleons and (2) nonrelativistic
extended nucleons.  We then generalize covariantly to obtain relativistic
equations of motion for extended nucleons, which can be written as
\begin{equation}\label{eqsmot}
M^*_i a_i^\mu = f_{{\rm s},i}^\mu + f_{{\rm v},i}^\mu + f_{{\rm s,ext},i}^\mu +
f_{{\rm v,ext},i}^\mu ~,
\end{equation}
where $M^*_i$ is the effective mass, $f_{{\rm s},i}^\mu$ is the scalar
self-force, $f_{{\rm v},i}^\mu$ is the vector self-force, $f_{{\rm
s,ext},i}^\mu$ is the scalar external force and $f_{{\rm v,ext},i}^\mu$ is the
vector external force.  These equations of motion are second-order, nonlinear,
integrodifferential equations with four dimensions per particle.
\bigskip

\section{Transverse momentum for soft nucleon-nucleon collisions}

To solve our equations of motion (\ref{eqsmot}) we use a fourth-order
Adams-Moulton predictor-corrector algorithm with adaptive step sizes.  The
integrations over proper time are done with a special error-minimizing
application of Lagrange's four-point (cubic) interpolation formulas.  We
present here results on the transverse momentum imparted in the soft collision
of two nucleons at laboratory momentum $p_{\rm lab}$ = 14.6, 30, 60, 100 and
200 G\eVcp   At three of these momenta substantial experimental data exist for
heavy-ion collisions\mycite{qm91,Ta89} and at the remaining two momenta
experimental data exist for proton-proton collisions.\mycite{Yo86}\ess  Results
calculated for the scattering angle, radiated energy and rapidity are described
in ref.~\cite{BN93b}.

\fig pt weiden/pt 238 {Calculated dependence of the transverse momentum upon
impact parameter for soft nucleon-nucleon collisions at five incident
laboratory momenta.}

As shown in fig.~\ref{pt}, for a given incident momentum, the transverse
momentum for soft reactions has a maximum value at a certain impact parameter
and decreases to zero for both head-on and distant collisions.  The maximum
transverse momentum increases slowly with increasing incident momentum in this
range, and the impact parameter at which this maximum occurs decreases.  For
ultrarelativistic collisions this impact parameter is approximately the
distance at which the transversely dominating static vector force for extended
nucleons\mycite{BN92,BN93} has its maximum.  At the other extreme of low
incident momentum, the opposing scalar and vector forces are of similar
magnitude and give rise for small impact parameter to the more complicated
behaviour of the double-dot--dashed curve in fig.~\ref{pt}.

The qualitative behaviour of our results can be understood in terms of the
nature of the external forces.  The repulsive vector force scales as the
Lorentz factor $\gamma$ in both the longitudinal and transverse directions,
whereas the attractive scalar force scales as $\gamma^2$ in the longitudinal
direction and as unity in the transverse direction.  This implies that the
vector force will dominate the transverse acceleration and the scalar force
will dominate the longitudinal acceleration.  For a given impact parameter the
scattering angle and transverse momentum will be essentially proportional to
the vector interaction strength~$\gv^2$, and the radiated energy will be
essentially proportional to $\gamma$ times the scalar interaction
strength~$\gs^2$.

As an initial test of the theory to describe the gross features of soft
nucleon-nucleon collisions, we compare in fig.~\ref{ptav} experimental and
theoretical values of the mean transverse momentum $<\!\!p_{\rm T}\!\!\!>$ as a
function of total centre-of-mass energy $\sqrt{s}$.  Because the classical
total cross section diverges, it is necessary when calculating such average
quantities to select a particular cutoff in impact parameter, or equivalently,
a value of the total cross section, for which purpose we have used\mycite{He90}
$\sigma_{\rm tot}$ = 40 mb.  Our results calculated with the values of the
constants listed in sect.~2 are given by the solid curve in fig.~\ref{ptav} and
are seen to be both lower in magnitude and to increase more rapidly with
$\sqrt{s}$ than the experimental values.\mycite{Al87,BL88}\ess However, the
theoretical curve is calculated for all values of rapidity corresponding to a
40 mb total cross section, whereas the experimental results are peak values in
the central region, where there is a short plateau in rapidity.  Furthermore,
the experimental values include high-$p_{\rm T}$ contributions arising from
hard collisions.  Because of the rough proportionality of $<\!p_{\rm T}\!\!>$
to the vector interaction strength $\gv^2$, we show with the dotted curve in
fig.~\ref{ptav} a simple estimate obtained by multiplying the solid curve by
the ratio of $\gv^2$ = 17.26 obtained by Bryan and Scott\mycite{BS69} from an
analysis of nucleon-nucleon scattering at laboratory kinetic energies between 0
and 350 M\eV to the value appearing in sect.~2.
\bigskip

\fig ptav weiden/ptavenp 238 {Comparison of experimental and theoretical values
of the mean transverse momentum $<\!\!p_{\rm T}\!\!\!>$ as a function of total
centre-of-mass energy $\sqrt{s}$.}

\section{Future directions}

{}From nucleon-nucleon scattering experiments we know that several additional
meson fields are important and must be included for a quantitative
description:\mycite{Ma89}
\bit
\item Isovector pseudoscalar ($\pi^+$, $\pi^-$, $\pi^0$)
\item Isovector scalar ($\delta^+$, $\delta^-$, $\delta^0$)
\item Isovector vector ($\rho^+$, $\rho^-$, $\rho^0$)
\item Neutral pseudoscalar ($\eta$)
\eit
The next step in the systematic development of the theory should be the
inclusion of these additional meson fields.  In addition, the effects of
quantum uncertainty on the equations of motion should be studied and included
if they are important.  This should be possible by use of techniques analogous
to those used by Moniz and Sharp for nonrelativistic quantum
electrodynamics.\mycite{MS77}

Once these two formal items are completed, the full version of the theory
should be used to calculate ultrarelativistic heavy-ion reactions that are
being studied experimentally.  Systematic comparisons between the calculated
and experimental results for such quantities as differential cross sections,
transverse momentum distributions, particle multiplicity distributions,
rapidity distributions and particle-particle correlations will enable us to
decide the usefulness of classical hadrodynamics for understanding these
phenomena.

This work was supported by the U.~S. Department of Energy.
\pagebreak

\end{document}